\documentclass[preprint,showpacs,preprintnumbers,amsmath,amssymb]{revtex4}

\usepackage{graphicx}
\usepackage{dcolumn}
\usepackage[dvips]{color}

\begin{document}

\title{Controlling Kerr nonlinearity with electric fields in asymmetric
double quantum-dots}

\author{Yi-Wen Jiang}
\email{ywjiang@sjtu.edu.cn}
\author{Ka-Di Zhu}%
\email{zhukadi@sjtu.edu.cn}
\affiliation{%
Department of Physics, Shanghai Jiao Tong University, 800 DongChuan
Road, Shanghai 200240, China
}%

\date{\today}

\begin{abstract}
 The control of Kerr nonlinearity with electric fields in an asymmetric double quantum-dot systems coupling
    with tunneling is investigated theoretically. It is found that,
    by proper tuning of two light beams and tunneling via a bias voltage, the Kerr
    nonlinearity can be enhanced and varied within a wide scale.
\end{abstract}

\maketitle

Over the past decades, there have been extensive researches on
nonlinear effects in multilevel systems, which mainly emphasized on
the enhancement of nonlinear optical properties
\cite{Yan,Nakajima,Matsko,Niu,Wang}. Recent evidences have shown
that possible applications would include frequency
conversion\cite{Merriam}, four-wave mixing\cite{Li} and harmonic
optical generation\cite{Hakuta}. In nonlinear optics, third-order
susceptibility plays an important role in information process. A
typical scheme for giant Kerr effect is proposed by Schmidt and
Imamoglu\cite{Schmidt}. The principle is a four-level atomic system
coupled by three fields: pump field, signal field and coupling
field, which consist an N configuration. This proposal had been
indirectly measured in the experiment \cite{Hau}. Harris and
Yamamoto \cite{Harris} also proposed a four-level atomic system
which existed a greatly enhanced Kerr nonlinearity, but vanishing
linear absorption. Recently, Sun \emph{et al.}\cite{Sun} studied the
enhanced Kerr nonlinearity in an asymmetric GaAs double quantum well
via Fano interference. Noticeably, most of Kerr effects in above
references are achieved by detuning of pump field or signal field.
In a recent report by  Sinclair and Korolkova \cite{Sinclair}, they
investigated Kerr effect considering the detuning of coupling field.

On the other hand, Villas-Boas \emph{et al.}\cite{Ulloa} proposed
a scheme of applying a tunneling voltage coupling two quantum-dots
(QDs)(see Fig.1). Yuan and Zhu \cite{Yuan} demonstrated
theoretically that there existed EIT in asymmetric double QD
systems coupling with a bias voltage. Comparing with
Ref.\cite{Nakajima}, we further investigate a four level system in
such asymmetric double quantum-dots system. However, different
from N-type in atomic systems, the upper two level are coupled by
tunneling, more like "\emph{n}-type". Because of the tunneling
between QDs via a bias voltage, the advantage of using
quantum-dots is that QDs allow direct control of its energy scales
and physical properties \cite{Ulloa}.

The quantum dot molecule (QDM) consisting of vertically stacked
pair of QDs coupled by tunneling via a bias voltage is shown in
Fig.1(a). One layer of highly $n^+$-doped GaAs and one layer of
undoped GaAs consist a QDM layer. Two layers are separated by 10
nm thick GaAs spacer \cite{Krenner}. The lower dot layer consists
of 8ML $Ga_{0.5}In_{0.5}As$, and the upper layer 7ML. By applying
a bias voltage between the $n$ contact and the Schottky gate, the
axial electric field can be tuned from 0 to $\sim 250 kV/cm$. We
model asymmetric double QDs consisting of two dots with different
band structure and the top two state $|3>, |4>$ are coupled by
tunneling (see Fig.1(b)). In this model, one could see left dot
(see Fig 1(b)) coupled with signal beam and the right dot coupled
with pump beam. The tunnel barrier in the double QDs can be
controlled by applying a bias voltage between two dots. With the
electromagnetic field, one exciton can be exited from the valence
to the conduction band of the right dot, which would then tunnel
to the left dot.

 We consider such a
Hamiltonian,
\begin{eqnarray}
H=\sum\varepsilon_j |j><j|+\hbar
\Omega_p(e^{-i\omega_pt}\mid3><1\mid+e^{i\omega_pt}\mid1><3\mid)\nonumber\\
+\hbar
\Omega_s(e^{-i\omega_st}\mid4><2\mid+e^{i\omega_st}\mid2><4\mid)+T_e(\mid3><4\mid+\mid4><3\mid).
\end{eqnarray}
where $\varepsilon_j$ is the energy of the state $|j> (j=1,2,3,4)$,
$T_e$ is the electron tunneling matrix element. $\Omega_p$ and
$\Omega_s$ are Rabi frequency for pump light and signal light,
respectively.
 Then we get,
\begin{eqnarray}
\dot{\rho_{24}}=-(i\omega_{24}+\gamma_{24})\rho_{24}-i\Omega_se^{-i\omega_st}(\rho_{44}-\rho_{22})
+iT_e\rho_{23},
\\\dot{\rho_{13}}=-(i\omega_{13}+\gamma_{13})\rho_{13}-i\Omega_pe^{-i\omega_pt}(\rho_{33}-\rho_{11})
+iT_e\rho_{14},
\\\dot{\rho_{34}}=-(i\omega_{34}+\gamma_{34})\rho_{34}-i(\Omega_pe^{i\omega_pt}\rho_{14}-\Omega_se^{-i\omega_st}\rho_{32})
-iT_e(\rho_{44}-\rho_{33}),
\\\dot{\rho_{12}}=-(i\omega_{12}+\gamma_{12})\rho_{12}-i(\Omega_pe^{-i\omega_pt}\rho_{32}-\Omega_se^{i\omega_st}\rho_{14}),
\\\dot{\rho_{14}}=-(i\omega_{14}+\gamma_{14})\rho_{14}-i(\Omega_pe^{-i\omega_pt}\rho_{34}-\Omega_se^{-i\omega_st}\rho_{12})
+iT_e\rho_{13},
\\\dot{\rho_{23}}=-(i\omega_{23}+\gamma_{23})\rho_{23}-i(\Omega_se^{-i\omega_st}\rho_{43}-\Omega_pe^{-i\omega_pt}\rho_{21})
+iT_e\rho_{24}.
\end{eqnarray}
with
$\omega_{24}=\varepsilon_2-\varepsilon_4,\omega_{13}=\varepsilon_1-\varepsilon_3,\omega_{34}=\varepsilon_3-\varepsilon_4,
\omega_{12}=\varepsilon_1-\varepsilon_2,\omega_{14}=\varepsilon_1-\varepsilon_4,\omega_{23}=\varepsilon_2-\varepsilon_3.$

Making the substitution,
\begin{eqnarray}
 \widetilde{\rho_{24}}=\rho_{24}e^{i\omega_st},
\widetilde{\rho_{13}}=\rho_{13}e^{i\omega_pt},
\widetilde{\rho_{34}}=\rho_{34},\nonumber\\
\widetilde{\rho_{12}}=\rho_{12}e^{i(\omega_p-\omega_s)t},
\widetilde{\rho_{14}}=\rho_{14}e^{i\omega_pt},
\widetilde{\rho_{23}}=\rho_{23}e^{i\omega_st}.\nonumber
\end{eqnarray}
The formulas can be rewritten as,
\begin{eqnarray}
\dot{\widetilde{\rho_{24}}}=-(i\Delta_{24}+\gamma_{24})\widetilde{\rho_{24}}-i\Omega_s(\widetilde{\rho_{44}}-\widetilde{\rho_{22}})
+iT_e\widetilde{\rho_{23}},
\\\dot{\widetilde{\rho_{13}}}=-(i\Delta_{13}+\gamma_{13})\widetilde{\rho_{13}}-i\Omega_p(\widetilde{\rho_{33}}-\widetilde{\rho_{11}})
+iT_e\widetilde{\rho_{14}},
\\\dot{\widetilde{\rho_{34}}}=-(i\omega_{34}+\gamma_{34})\widetilde{\rho_{34}}-i(\Omega_p\widetilde{\rho_{14}}-\Omega_s\widetilde{\rho_{23}}^*)
-iT_e(\widetilde{\rho_{44}}-\widetilde{\rho_{33}}),
\\\dot{\widetilde{\rho_{12}}}=-(i\Delta_{12}+\gamma_{12})\widetilde{\rho_{12}}-i(\Omega_p\widetilde{\rho_{23}}^*-\Omega_s\widetilde{\rho_{14}}),
\\\dot{\widetilde{\rho_{14}}}=-(i\Delta_{14}+\gamma_{14})\widetilde{\rho_{14}}-i(\Omega_p\widetilde{\rho_{34}}-\Omega_s\widetilde{\rho_{12}})
+iT_e\widetilde{\rho_{13}},
\\\dot{\widetilde{\rho_{23}}}=-(i\Delta_{23}+\gamma_{23})\widetilde{\rho_{23}}-i(\Omega_s\widetilde{\rho_{34}}^*-\Omega_p\widetilde{\rho_{12}}^*)
+iT_e\widetilde{\rho_{24}}.
\end{eqnarray}
where $\gamma_{13},\gamma_{24},\gamma_{34}$ represent decay rates
for $\rho_{13}, \rho_{24}, \rho_{34}$.

Under single electron approximation, the double quantum-dot system
is initially in the ground state $|2>$, so we assume that
$\widetilde{\rho}_{22}^{(0)}=1,\widetilde{\rho}_{11}^{(0)}=\widetilde{\rho}_{33}^{(0)}=\widetilde{\rho}_{44}^{(0)}=0.$
And at the initial, the pump light and the voltage are strong, so
$\widetilde{\rho}_{13}^{(0)}=0,\widetilde{\rho}_{34}^{(0)}=0$.

This set of equations can be solved,
\begin{eqnarray}
|M|=(i\Delta_{24}+\gamma_{24})(i(\Delta_{13}-\Delta_{24}-\omega_{34})+\gamma_{12})(i(\Delta_{13}+\omega_{34})+\gamma_{14})(i(\Delta_{24}-\omega_{34})+\gamma_{23})
\nonumber\\-(i\Delta_{24}+\gamma_{24})(i(\Delta_{13}+\omega_{34})+\gamma_{14})\Omega_p^2-T_e^2
\Omega_s^2,
\\ \widetilde{\rho}_{24}=\frac{i\Omega_s}{|M|}[-(i(\Delta_{13}-\Delta_{24}-\omega_{34})+\gamma_{12})(i(\Delta_{13}+\omega_{34})+\gamma_{14})(i(\Delta_{24}-\omega_{34})+\gamma_{23})
\nonumber\\-(i(\Delta_{24}-\omega_{34})+\gamma_{23})\Omega_s^2+(i(\Delta_{13}+\omega_{34})+\gamma_{14})\Omega_p^2].
\end{eqnarray}
We then obtain
\begin{eqnarray}
\chi^{(3)}=\frac{N
{\rho}_{24}}{3E_p^2E_s}=\frac{N\mu^4}{3\hbar^3}\chi^{(3)}_{eff},
\\\chi^{(3)}_{eff}=\frac{i(i(\Delta_{13}+\omega_{34})+\gamma_{14})}{\alpha+i\beta}
=\frac{-(\Delta_{13}+\omega_{34})\alpha+i\beta(\Delta_{13}+\omega_{34})+i\alpha\gamma_{14}+\beta\gamma_{14}}{\alpha^2+\beta^2}
\end{eqnarray}
with
\begin{eqnarray}
 \alpha=(\Delta_{13}-\Delta_{24}-\omega_{34})(\Delta_{13}+\omega_{34})(\Delta_{24}-\omega_{34})\Delta_{24}-\Delta_{24}(\Delta_{13}-\Delta_{24}-\omega_{34})\gamma_{14}\gamma_{23}\nonumber\\
-(\Delta_{13}+\omega_{34})\Delta_{24}\gamma_{12}\gamma_{23}
-(\Delta_{24}-\omega_{34})\Delta_{24}\gamma_{12}\gamma_{14}-(\Delta_{13}-\Delta_{24}-\omega_{34})(\Delta_{13}+\omega_{34})\gamma_{23}\gamma_{24}
\nonumber\\-(\Delta_{13}-\Delta_{24}-\omega_{34})(\Delta_{24}-\omega_{34})\gamma_{14}\gamma_{24}-(\Delta_{13}+\omega_{34})(\Delta_{24}-\omega_{34})\gamma_{12}\gamma_{24}
+\gamma_{12}\gamma_{14}\gamma_{23}\gamma_{24}\nonumber
\\+(\Delta_{24}(\Delta_{13}+\omega_{34})-\gamma_{14}\gamma_{24})\Omega_p^2-T_e^2\Omega_s^2,
\\ \beta=-(\Delta_{13}+\omega_{34})(\Delta_{13}-\Delta_{24}-\omega_{34})(\Delta_{24}-\omega_{34})\gamma_{23}-(\Delta_{24}-\omega_{34})\Delta_{24}(\Delta_{13}-\Delta_{24}-\omega_{34})\gamma_{14}
\nonumber\\-(\Delta_{13}+\omega_{34})(\Delta_{24}-\omega_{34})\Delta_{24}\gamma_{12}+\Delta_{24}\gamma_{12}\gamma_{14}\gamma_{23}
-(\Delta_{13}-\Delta_{24}-\omega_{34})(\Delta_{13}+\omega_{34})(\Delta_{24}-\omega_{34})\gamma_{24}
\nonumber\\+(\Delta_{13}-\Delta_{24}-\omega_{34})\gamma_{23}\gamma_{14}\gamma_{24}+(\Delta_{13}+\omega_{34})\gamma_{23}\gamma_{12}\gamma_{24}
+(\Delta_{24}-\omega_{34})\gamma_{12}\gamma_{14}\gamma_{24}
\nonumber\\-(\Delta_{24}\gamma_{14}+(\Delta_{13}+\omega_{34})\gamma_{24})\Omega_p^2.
\end{eqnarray}

The real and imaginary parts of $\chi^{(3)}_{eff}$,
representatively, yield the Kerr nonlinearity and nonlinear
absorption. In the model we presented before, the nonlinear
properties can be controlled through the intensity $\Omega_p$ and
$\Omega_s$, detunings $\Delta_{13}$ and $\Delta_{24}$, and
tunneling $T_e$ via a bias voltage. Since $\Omega_s$ is much
smaller than $\Omega_p$, we focus on changing of $\Omega_p$. Fig.2
shows the normalized Kerr nonlinearity index and the nonlinear
absorption as a function of the detuning $\Delta_{24}$ with
tunneling $T_e$ equal to $10\gamma_{24}$. We choose
$\gamma_{12}=\gamma_{14}=\gamma_{23}=0$, because they are much
smaller than decay rate $\gamma_{24}$, which is usually the case.
When the detuning $\Delta_{13}=0$, which means both signal and
pump light are on resonance, Kerr nonlinearity would increase with
increasing $\Omega_p$. Taking Fig.2(a) as an example, when
detuning $\Delta_{24}/\gamma_{24}=0.7$, the nonlinear absorption
is approximately zero, while Kerr nonlinearity reaches its
maximum. This result indicates that Kerr nonlinearity can be
enhanced, while the nonlinear absorption is cancelled. Comparing
Fig.2(a) with (c), if $\Omega_p$ is fixed, Kerr nonlinearity is
larger under off-resonant condition.  In order to obtain the
maximal Kerr nonlinearity with a certain bias voltage,
experimentally, one should choose both large $\Omega_p$ and
$\Delta_{24}$.

The Kerr nonlinearity as a function of tunneling $T_e$ is plotted
in Fig.3. Here are parameters we choose: the quantum-dot has a
base length of $9nm$ and a height of $3.5nm$and an area density of
about $4\times10^{-10} cm^2$ \cite{Jiang}. The interband
transition energy for each dot are $1.1 eV$ and $1.3 eV$
respectively. The bandwidth of signal laser is $100Gbits/s$, which
corresponding to $\hbar \gamma_{24}=66\mu eV$. $\hbar \Omega_p$
and $\hbar \Omega_s$ are $1 eV$ and $0.1 eV$, respectively, since
pump beam is strong. When bias voltage is zero, the whole system
turns to be EIT in QD systems, which we call Phonon Induced
Transparency\cite{Jiang}. Four lines give the off-resonance
condition of both pump light and signal light. Comparing
$\Delta=-3$ with $\Delta=3$, and $\Delta=-6$ with $\Delta=6$, if
detuning is positive, Kerr nonlinearity begins from negative
value. Conversely, Kerr nonlinearity would be positive at the
beginning if detuning is negative. All four curves indicate that
Kerr nonlinearities have a maximum point and a minimum point. When
detuning is positive, Kerr nonlinearity first increases to maximum
point, then decreases to minimum point as the tunneling $T_e$
increases. Eq.(17) exhibits that the contributions to the
third-order susceptibility consists of two part: the first is the
influence of the pump light; the second part is the transit
between level $|1>$ and $|4>$ which is negligible. So in Fig.3,
the off-resonant line shape is indeed determined by pump field.
When a light beam travels in a homogeneous Kerr medium, positive
nonlinearity ($>0$) always leads to self-focusing, while negative
nonlinearity ($<0$) leads to self-defocusing. This figure shows
that by properly controlling the bias voltage, the signal light
can be either self-focusing, or self-defocusing.

In conclusion, we have theoretically investigated the enhancement of
Kerr nonlinearity in asymmetric double QDs. The results show that by
proper tuning of the bias voltage, giant Kerr nonlinearities can be
realized while cancelling nonlinear absorption. This structure also
indicates that a wide range of Kerr nonlinearity from negative to
positive can be simply adjusted via a bias voltage, which would
result in wide control of signal beam.

The part of this work was supported by National Natural Science
Foundation of China (No.10774101) and the National Ministry of
Education Program for Training Ph.D..

\newpage
\centerline{\large{\bf Figure Captions}}

Fig.1.(a) Scheme of the setup. Signal light and pump light transit
through QDs. $V_B$ is a bias voltage. (b) Energy level scheme for an
asymmetric double QD system. $|3>$ and $|4>$ are coupled by the
tunneling $T_e$.

Fig.2 The real part and imaginary part of the linear optical
susceptibility as a function of the detuning
$\Delta_{24}/\gamma_{24}$. The solid lines are Kerr nonlinearities,
and the dashed lines are nonlinear absorption. Parameters are chosen
as: $\omega_{34}/\gamma_{24}=1,\Omega_s/\gamma_{24}=0.1,
T_e/\gamma_{24}=10,\gamma_{12}=\gamma_{14}=\gamma_{23}=0$.

Fig.3 The Kerr coefficient as a function of tunneling
$T_e/\gamma_{24}$. On-resonance and near-resonance conditions are
considered. Parameters are chosen as: $\omega_{34}/\gamma_{24}=1,
\gamma_{12}=\gamma_{14}=\gamma_{23}=0$. $\hbar \Omega_p$ and
$\hbar \Omega_s$ are $1 eV$ and $0.1 eV$,respectively.

 \clearpage
\begin{figure}
\includegraphics{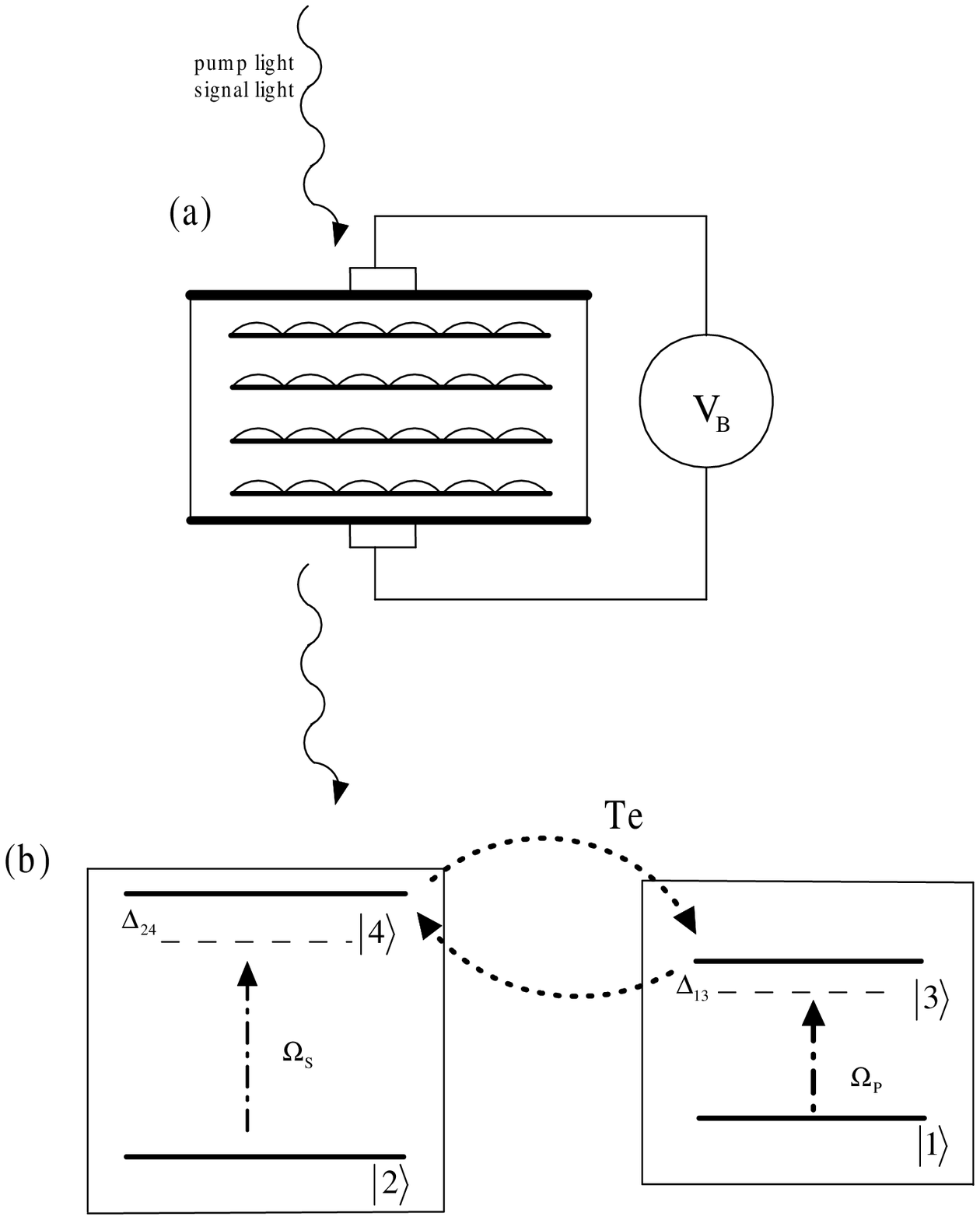}
\caption{}
\end{figure}

\clearpage
\begin{figure}
\includegraphics{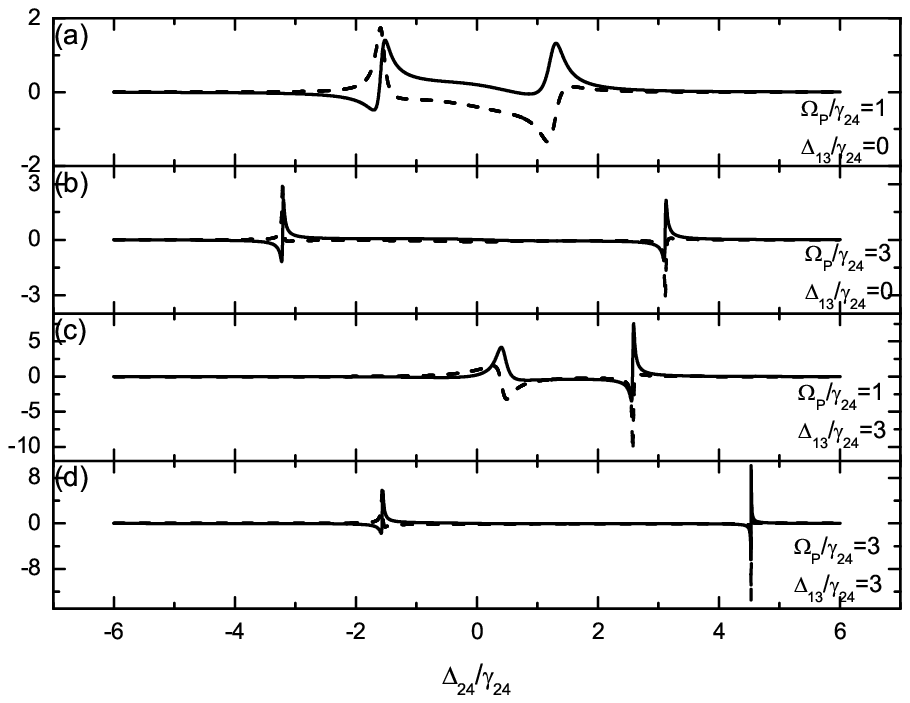}
\caption{}
\end{figure}

\clearpage
\begin{figure}

\includegraphics{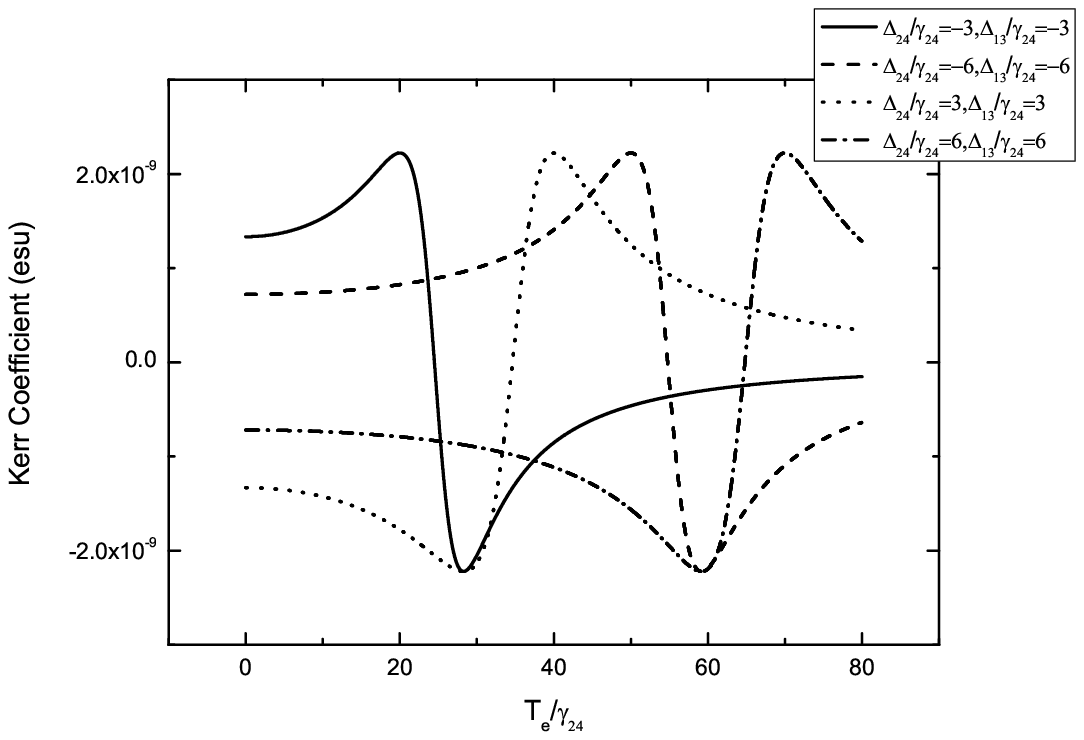}
\caption{}
\end{figure}

\bibliography{apssamp}

\end{document}